\documentclass[preprint,12pt]{elsarticle}

\usepackage{amssymb}

\usepackage{lineno}

\journal{Nuclear Instruments and Methods A}

\begin{document}

\begin{frontmatter}



\title{Diamonds as timing detectors for MIP: The HADES proton{}-beam
monitor and start detectors}


\author[pie]{J. Pietraszko \corref{cor1}}
\author[fab]{L. Fabbietti}
\author[pie]{W. Koenig}
\author{\\(for the HADES collaboration)}

\address[pie]{GSI Helmholtz Centre for Heavy Ion Research
GmbH Planckstrasse 1, \\D-64291 Darmstadt, GERMANY}
\address[fab]{Excellence Cluster 'Universe', Technische Universit\"{a}t M\"{u}nchen,
Boltzmannstr. 2, D-85748 Garching Germany. }
\cortext[cor1]{Corresponding author's e-mail address:
j.pietraszko@gsi.de}

\begin{abstract}
This paper gives an overview of a recent development of measuring
time of flight of minimum-ionizing particles (MIP) with
mono-crystalline diamond detectors. The application in the HADES
spectrometer as well as test results obtained with proton beams
are discussed.
\end{abstract}


\end{frontmatter}



\section{Introduction} Diamond detectors are well known for their
radiation hardness and high drift speeds of both electrons and
holes, making them ideal not only as timing detectors placed in
the beam\,\cite{diamonds1, diamonds2} but also as luminosity
monitors\,\cite{cms_lum}. However, due to the large effective
energy needed to create electron-hole pairs (13\,eV) the charge
created by minimum-ionizing particles (MIP) traversing the diamond
is marginal (8000 pairs for a 300\,$\mu$m thick
diamond)\,\cite{pomorski_1}. Taking into account that the
poly-crystalline texture of CVD-diamond material leads to losses
in charge collection these sensors can not be used for the
detection of MIP because signals are too small to be registered by
state-of-the-art electronics. Instead, the recently developed
technology of producing a mono-crystalline diamond material, which
is almost free of structural defects and chemical impurities and
thus provides very high charge collection efficiency, allows for
building detectors for MIP based on diamond material. \\In the
following a dedicated, low-noise readout scheme is described as
well as its application as a start detector for proton beams in
the HADES experiment\,\cite{hades_nim}.

\section{Diamond Readout} 
\begin{figure}[thb]
\begin{minipage}[t]{0.48\textwidth}
      \begin{picture}(160,160)(0,0)
      \put(20,0){ \includegraphics[width=0.8\textwidth,height=0.7\textwidth]
          {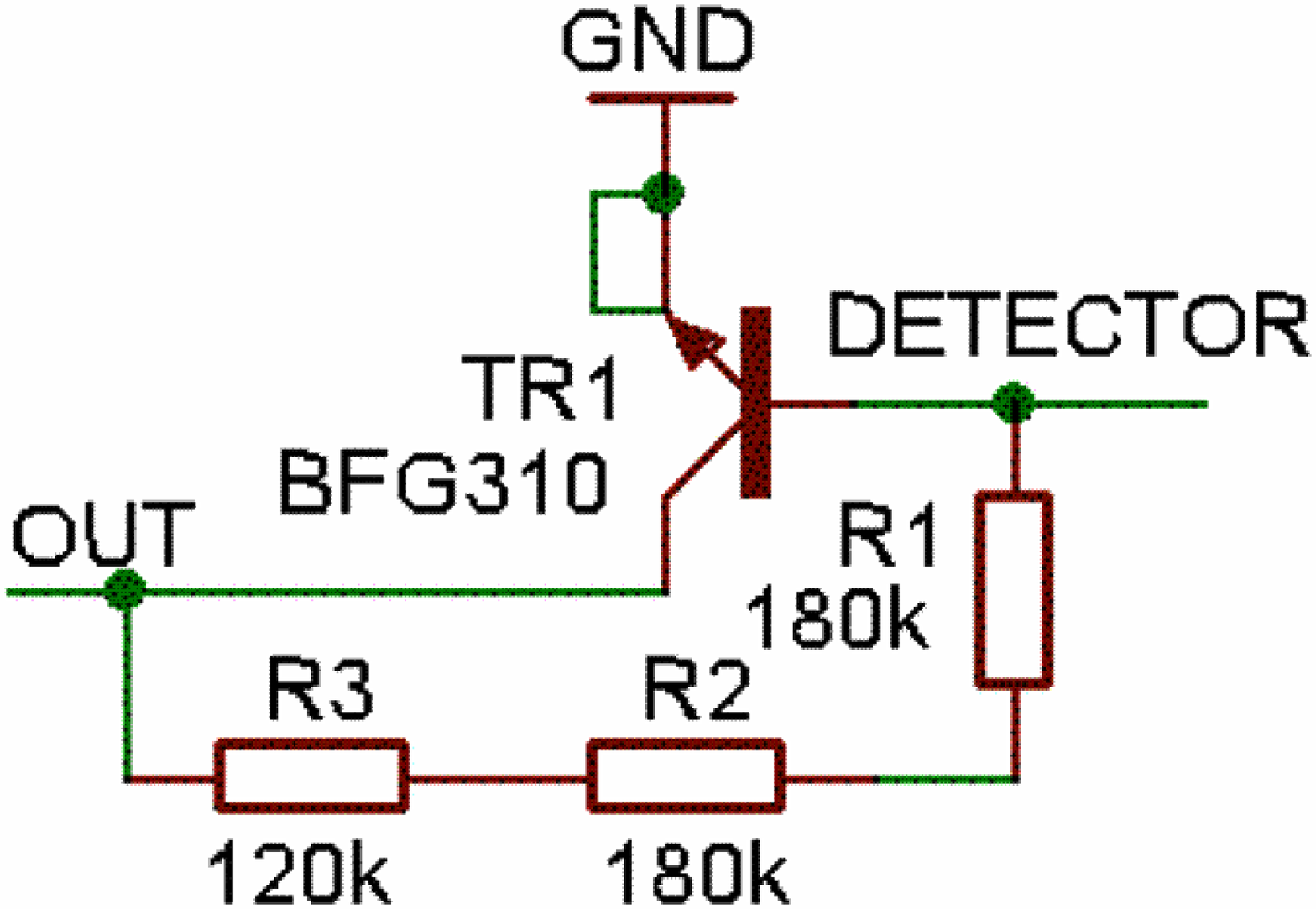}}
      \end{picture}
 \caption{Schematics of the front-end amplifier.}
\label{fig:electronics}
\end{minipage}
\hspace{\fill}
\begin{minipage}[t]{0.48\textwidth}
      \begin{picture}(160,160)(0,0)
      \put(0,0){ \includegraphics[width=0.98\textwidth,height=.8\textwidth]
          {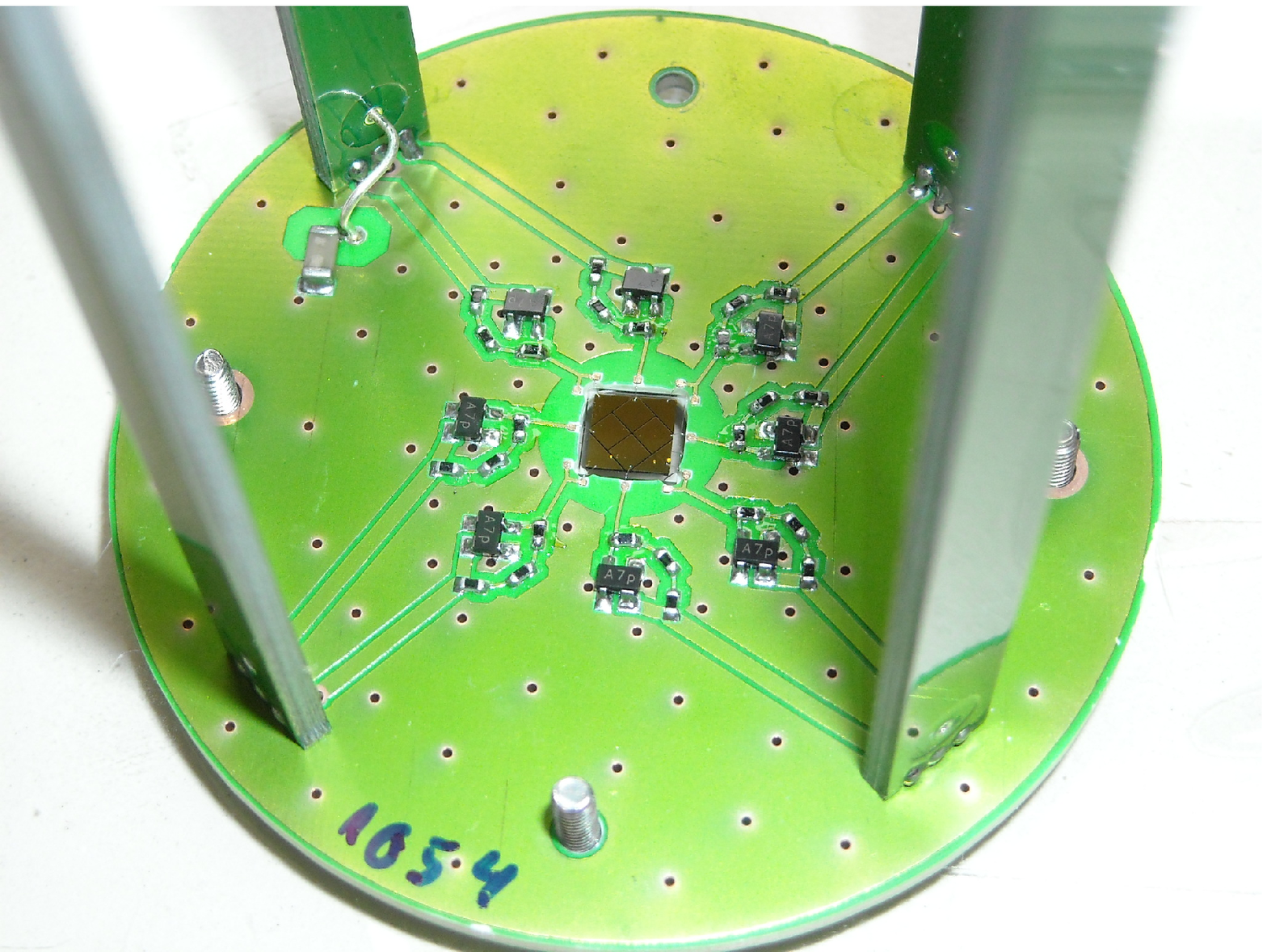}}
      \end{picture}
\caption{PC board (${\phi}$\,=\,50\,mm) with the diamond
(4.7$\times$~4.7\,mm$^2$) in the centre surrounded by
8\,amplifiers.} \label{fig:pcb}
\end{minipage}
\end{figure}

Mono-crystalline diamonds with two
different detector sizes of 3.5$\times$3.5\,mm$^2$ (4\,pixel) and
4.7$\times$4.7\,mm$^2$ (8\,pixel) with thicknesses of 300\,$\mu$m
and 500\,$\mu$m, respectively, were used during the test. For both
our detectors a capacitance of a single segment was 0.25\,pF only,
which was defined by the geometry of the detectors. The front-end
electronics has been build using RF transistors with low input
capacitance (0.2\,pF) which were placed close to the diamond (see
Fig.\,\ref{fig:electronics},\,\ref{fig:pcb}). The bias current was
reduced to an extremely small value leading to a relatively large
input impedance of about 2\,k${\Omega}$.
This results in an integration of the primary charge signal. Final
shaping is done by an external booster amplifier resulting in
rise-times ($10\% - 90\%$) of 1.2\,ns (300\,$\mu$m) and 1.35\,ns
(500\,$\mu$m), respectively. The booster amplifier contains a
56\,${\Omega}$ pullup resistor to 5\,V at the input to provide the
bias voltage as well as proper termination of the signal line. It
should be mentioned that for such a front-end design it is
necessary to keep the stray capacitances at a
minimum\,\cite{norhdia_2008}. In our case this was realized by the
bias current which is provided via 3\,resistors in series in order
to reduce their capacitive coupling below 0.1\,pF. The power
consumption of a single amplifier amounts to slightly less than
5\,mW at a 5\,V supply allowing for operation in vacuum which is
an important issue for the foreseen application as a beam detector.\\
The base plate containing a diamond (4.7$\times$4.7\,mm$^2$)
surrounded by 8\,amplifiers is shown in Fig.\,\ref{fig:pcb}. The
PC board material is Rogers 4003C TM utilizing a low dielectricity
constant of 3.4.
\section{Results} The diamond detectors were exposed to proton beams
with kinetic energies from 1.2\,GeV to 3.5\,GeV and rates of up to
3$\times$10$^{6}$/s/10$mm^2$. Several detector samples with
different metallizations were tested to achieve stable operation
of the detector at high beam intensities. As a result the
high-rate capability could be reached only after applying a
metallization procedure which includes passivation in an oxygen
plasma, "baking" at 500$^o$C in an Ar atmosphere and depositing a
50\,nm Cr layer followed by a 150\,nm Au layer. Otherwise, a
constantly increasing leakage current appeared at intensities
above several 10$^{5}$/s which finally resulted in a rapid
discharge.
\begin{figure}[htb]
\begin{minipage}[t]{0.48\textwidth}
      \begin{picture}(160,190)(0,0)
      \put(-5,-19){ \includegraphics[width=0.98\textwidth,height=1.05\textwidth]
          {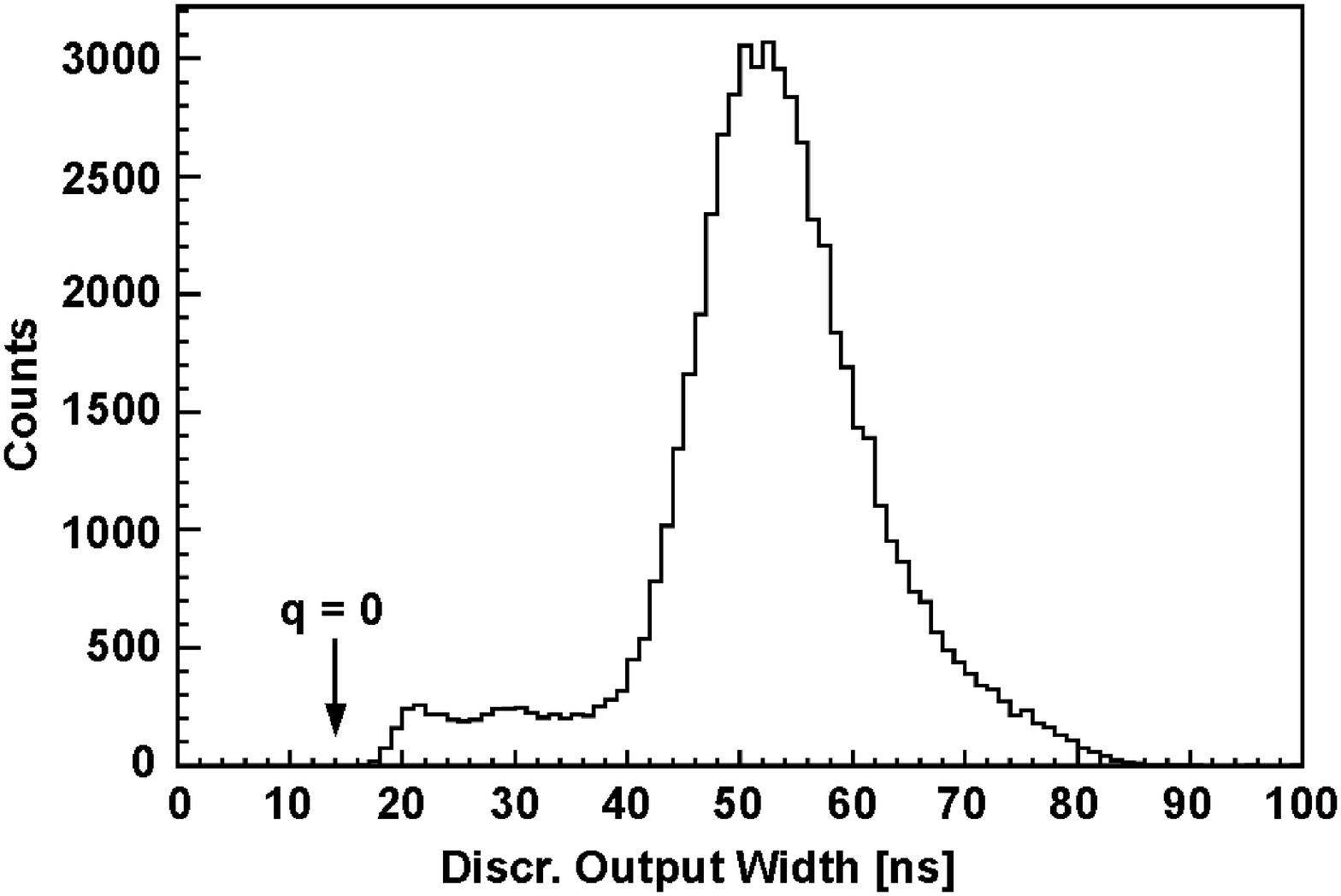}}
      \end{picture}
\caption{Pulse-charge distribution measured via charge to
pulse-width conversion for 1.8\,GeV protons traversing a
mono-crystalline diamond of 500\,$\mu$m thickness.}
\label{fig:pulse}
\end{minipage}
\hspace{\fill}
\begin{minipage}[t]{0.48\textwidth}
      \begin{picture}(160,190)(0,0)
      \put(0,-15){ \includegraphics[width=1.05\textwidth,height=.980\textwidth]
          {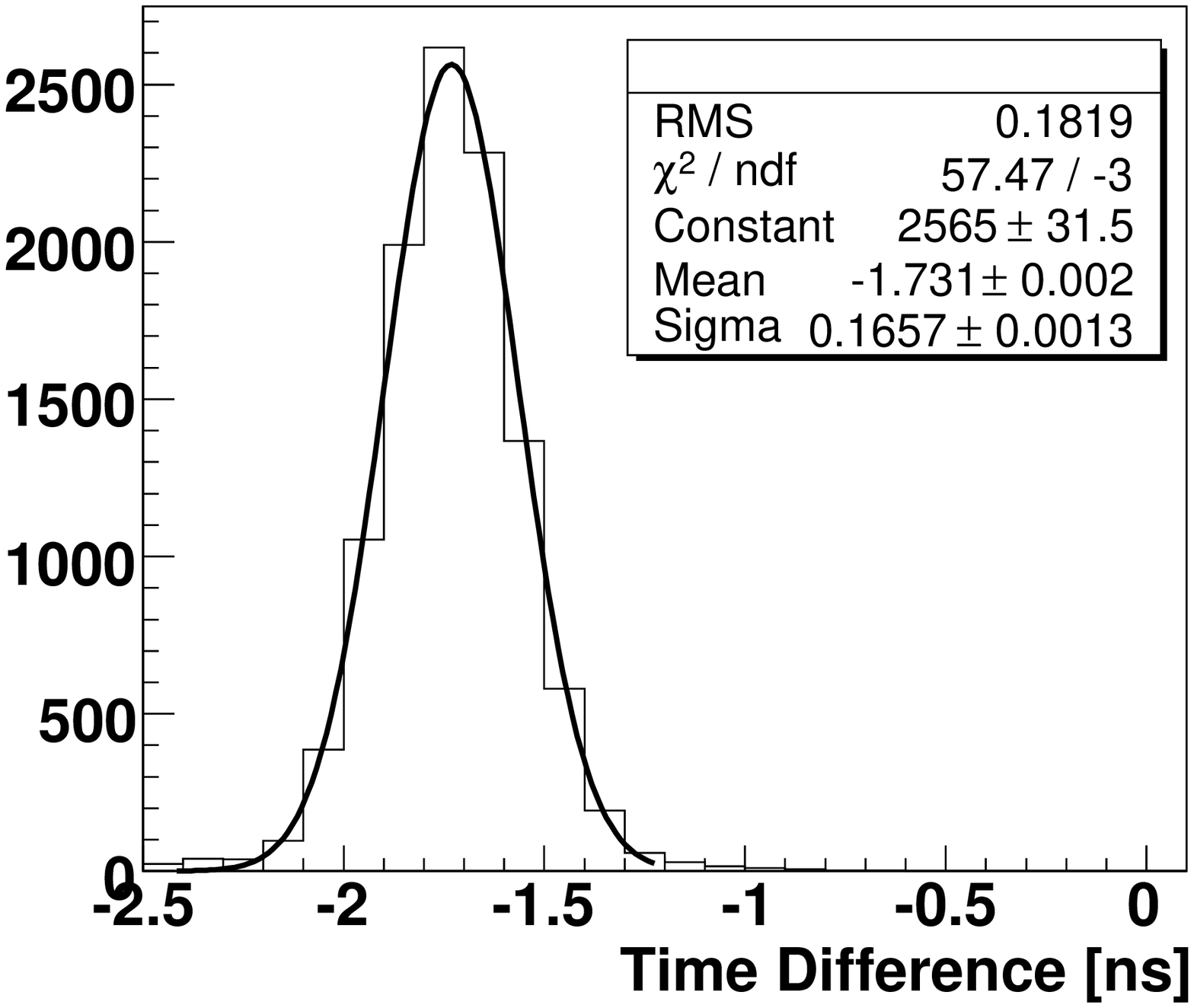}}
      \end{picture}
\caption{Time resolution for 1.8\,GeV protons between two
mono{}-crystalline diamonds without any correction.}
\label{fig:time_res}
\end{minipage}
\end{figure}
A data acquisition system of the test setup was built based on the
TRB\,\cite{hades_trb1}, a stand-alone DAQ board, which employs the
HPTDS\,\cite{hptds_1} chips for time measurement. The analog
signal from the detector, after amplification, was sent to the
signal discrimination circuit with the signal integration
functionality. The integrated signal was converted into pulse
width of the timing output of the discriminator which allowed for
pulse charge measurement. Fig.\,\ref{fig:pulse} shows the signal
charge distribution of a single segment measured for 1.8\,GeV
protons (MIP). Zero charge results in a minimum width of 14\,ns
(arrow in Fig.\,\ref{fig:pulse}). The visible tail at small signal
charges (small signal widths) is dominated by cases where the beam
particle hits the detector at the edge of the segment and in this
case charge is shared between neighbored segments of the diamond.

To determine the time resolution of the detectors, two diamonds
were put into the proton beam with a distance of 10\,cm between
them. As shown in Fig.\,\ref{fig:time_res}, the time resolution
measured with leading edge discriminators between two segments of
diamond is of the order of 165\,ps. This gives a single segment
time resolution of 117\,ps. Based on the signal to RMS{}-noise
ratio of about 23 (300\,${\mu}$m) and 27 (500\,${\mu}$m) one
expects a intrinsic time resolution of about 65\,ps for both
detector thicknesses which clearly gives room for further
improvements.

The detection efficiency of the device was measured with a well
focused proton beam centered in the middle of the diamond
detector. The diameter of the beam spot at the focal point (center
of the diamond) was below 2\,mm and the halo of proton beams is
typically below 1$\%$. As a reference, a plastic, segmented
detector, located 7\,m downstream, where the beam was defocused
(beam spot diameter about 30\,mm), was used. By comparing the
rates measured in the reference detector with the rates seen by
the diamond detector, the diamond detection efficiency was
determined to be ${\geq}$95\%.
\section{Outlook} Due to continuous progress in the development
of low-noise transistors, the signal to RMS{}-noise ratio can be
improved by nearly 50$\%$ at 10$\%$ shorter rise-time based on up
to date SiGe:C technology. In particular, tiny housings provide
less stray capacitance (E.g. BFR705L3RH). The power consumption
per channel can be reduced to below 2.5\,mW and the total area of
all components of such an amplifier amounts to only 2\,mm$^2$.
First results with a Sr$^{90}$ source confirmed these
improvements, but no in-beam tests were performed so far.
\section{Acknowledgements} For the preparation of the
detectors, metallization and bonding of the diamonds we highly
appreciate the support of E. Berdemann, M. Tr\"ager et al., GSI
Detector Laboratory and A. H\"ubner et al., GSI Target
Laboratory.\\


\textbf{References}
\begin{thebibliography}{00}
\bibitem{diamonds1} W. Adam et al., The RD42 Collaboration, NIM A 511 (2003) 124
\bibitem{diamonds2} C. Mer et al., Diamond and Related Materials, 13 (2004) 791
\bibitem{cms_lum} E. Bartz et al., JINST 4 P04015 2009
\bibitem{hades_nim}G. Agakichiev et al. (HADES), Eur. Phys. J. A 41 (2009) 243
\bibitem{norhdia_2008}The 4th NoRHDia Workshop, GSI, Darmstadt June 8 - June 10,
2008\\
http://www-norhdia.gsi.de/talks/4th/W\_Koenig.pdf,\\
http://www-norhdia.gsi.de/talks/4th/A\_Schuettauf.pdf
\bibitem{hades_trb1}M. Traxler, 128 channel high resolution TDC with integrated
DAQ-system, GSI Scientific Report (2005) 281
\bibitem{hptds_1}HPTDC, J. Christiansen, Digital Microelec. Group, CERN
\bibitem{pomorski_1} Single crystal cvd diamond detectors for hadron physics , M.
Pomorski et al, Proceedings of the 9th Conference Astroparticle,
particle and space physics, detectors and medical physics
applications, Villa Olmo, Como, Italy, 17-21 October 2005 , p 92,
DOI No: 10.1142/9789812773678\_0016
\end{thebibliography}
\end{document}